\documentclass[prl,twocolumn,showpacs,preprintnumbers,amsmath,amssymb]{revtex4}

\usepackage{graphicx}
\usepackage{dcolumn}
\usepackage{bm}

\begin{document}


\title{Precision lifetime measurements of a single trapped ion with ultrafast laser pulses}

\author{D. L. Moehring}
 \email{dmoehrin@umich.edu}
\author{B. B. Blinov}
\author{D. W. Gidley}
\author{R. N. Kohn Jr.}
\author{M. J. Madsen}
\author{T. D. Sanderson}
\author{R. S. Vallery}
\author{C. Monroe}
\affiliation{FOCUS Center and Department of Physics, University of Michigan, Ann Arbor, Michigan 48109-1120}

\date{\today}

\begin{abstract}
We report precision measurements of the excited state lifetime of the $5p$ $^2P_{1/2}$ and $5p$ $^2P_{3/2}$ levels of a single trapped Cd$^+$ ion.  The ion is excited with picosecond laser pulses from a mode-locked laser and the distribution of arrival times of spontaneously emitted photons is recorded.  The resulting lifetimes are 3.148 $\pm$ 0.011 ns and 2.647 $\pm$ 0.010 ns for $^2P_{1/2}$ and $^2P_{3/2}$ respectively.  With a total uncertainty of under 0.4\%, these are among the most precise measurements of any atomic state lifetimes to date.
\end{abstract}


\maketitle

Precise measurements of atomic data are of great interest throughout many fields of science.  Lifetime measurements are of particular importance to the interpretation of measurements of atomic parity nonconservation \cite{wood:1997}, tests of QED and atomic structure theory \cite{curtis}, and even astrophysical applications \cite{li:2000}.  Because of this, new and more accurate ways of measuring excited state lifetimes are constantly being investigated.  Previous methods include time-correlated single photon techniques \cite{young:1994, devoe:1994, devoe:1996, pinnington:1994, aubin:2003, xu:2004}, beam-foil experiments \cite{pinnington:1994}, fast beam measurements \cite{gaupp:1982, rafac:1999}, electron-photon delayed coincidence techniques \cite{imhof:1971, shaw:1975}, luminescent decay \cite{lefers:2002, zinner:2003}, linewidth measurements \cite{oates:1996}, photoassociative spectroscopy \cite{mcalexander:1995}, and quantum jump methods \cite{peik:1994}.  

Here we report excited state lifetime measurements using a time-correlated single photon counting technique on a single atom.  This method, designed especially to eliminate common systematic errors, involves selective excitation of a single trapped ion to a particular excited state (lifetime of order nanoseconds) by an ultrafast laser pulse (duration of order picoseconds).  Arrival of the spontaneously-emitted photon from the ion is correlated in time with the excitation pulse, and the excited state lifetime is extracted from the distribution of time delays from many such events.  

By performing the experiment on a single trapped ion, we are able to eliminate prevalent systematic errors, such as: pulse pileup that causes multiple photons to be collected within the time resolution of the detector, radiation trapping or the absorption and re-emission of radiation by neighboring atoms, atoms disappearing from view before decaying, and subradiance or superradiance arising from coherent interactions with nearby atoms.  By using ultrafast laser pulses, we eliminate potential effects from applied light during the measurement interval.  

With this setup, at most one photon can be emitted following an excitation pulse.  While this feature is instrumental in eliminating the above systematic errors, it would appear that this signal would require large integration times
for reasonable statistical uncertainties.  However, with a lifetime of only a few nanoseconds, millions of such excitations can be performed each second, thus potentially allowing sufficient data for a statistical error of under 0.1\% to be collected in a matter of minutes \cite{devoe:1994}.

\begin{figure}
\includegraphics[width=1.0\columnwidth,keepaspectratio]{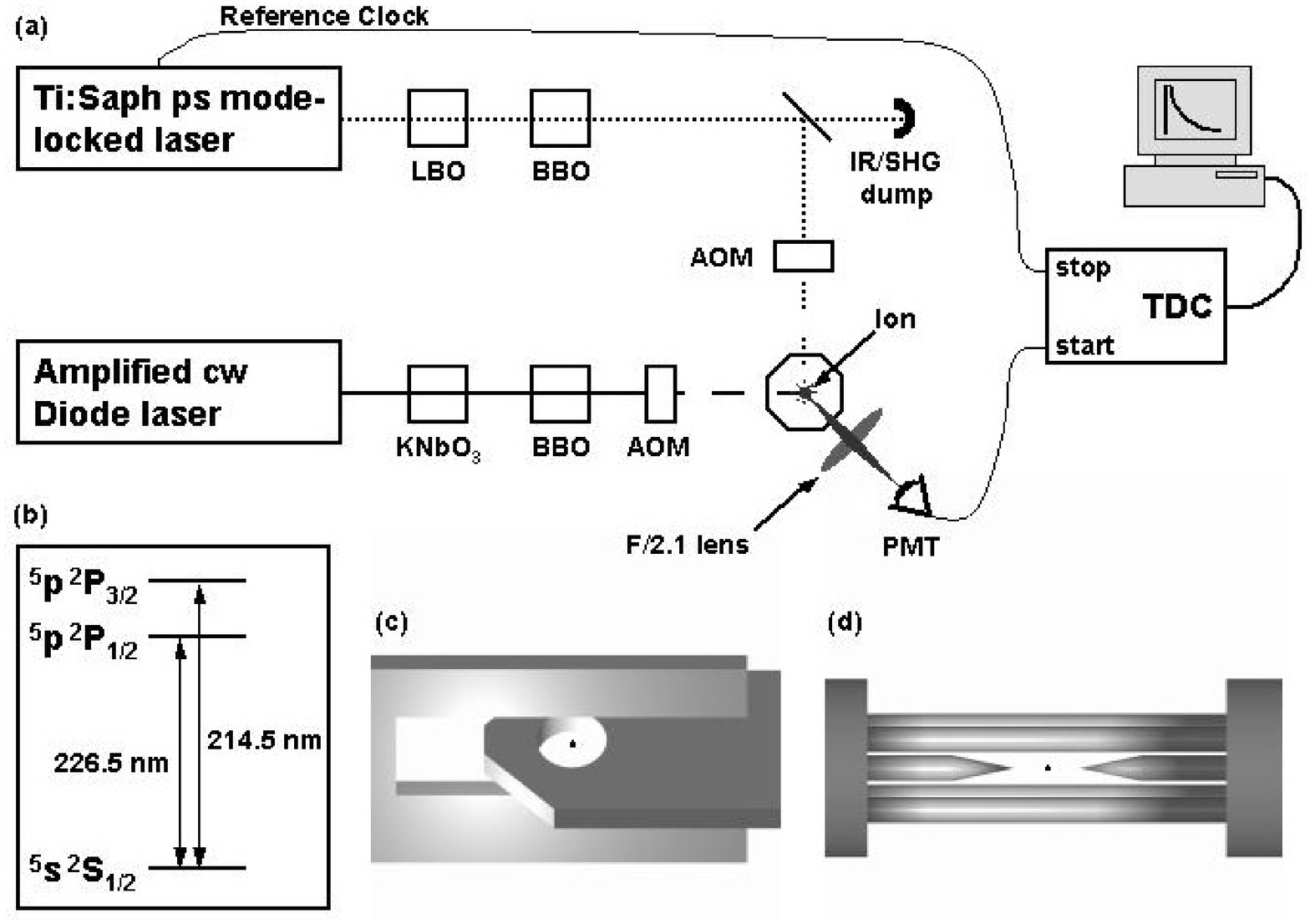}
\caption{\label{fig:apparatus}The experimental apparatus.  (a) A picosecond mode locked Ti:Saph laser is tuned to four times the resonant wavelength for either the $5p$ $^2P_{1/2}$ or the $5p$ $^2P_{3/2}$ level of Cd$^+$.  Each pulse is then frequency-quadrupled through non-linear crystals, filtered from the fundamental and second harmonic, and directed to the ion.  An amplified cw diode laser is also frequency quadrupled and tuned just red of the $^2P_{3/2}$ transition for Doppler cooling of the ion within the trap.  Acousto-optic modulators (AOM) are used to switch on and off the lasers as described in the text.  Photons emitted from the ion are collected by an $f/2.1$ imaging lens and directed toward a photon-counting photo multiplier tube (PMT).  The output of the PMT provides the start pulse for the time to digital converter (TDC), whereas the stop pulse is provided by the reference clock of the mode-locked laser.  (b) The relevant energy levels of Cd$^+$.  (c) An asymmetric quadrupole trap.  (d) A linear trap.}
\end{figure}

A diagram of the experimental apparatus is shown in Fig.~\ref{fig:apparatus}.  Individual cadmium ions are trapped and isolated in one of two rf quadrupole traps.  First, the experiment is conducted using an asymmetric quadrupole trap of characteristic size $\sim$0.7 mm \cite{jefferts:1995} [Fig. ~\ref{fig:apparatus}(c)].  The entire experiment is then repeated in a linear trap with rod spacings of 0.5 mm and an endcap spacing of 2.6 mm [Fig. ~\ref{fig:apparatus}(d)].  Both traps have secular trapping frequencies on the order of $\omega/2\pi\sim0.1-1.0$ MHz.  

Two types of laser radiation are incident on the ion: pulsed and continuous wave (cw) lasers.  The pulsed light is from a picosecond mode-locked Ti:Sapphire laser whose center frequency is resonantly tuned to provide excitation to one of the $^2P$ states [Fig. ~\ref{fig:apparatus}(b)].  For excitation to the $5p$ $^2P_{1/2}$ ($5p$ $^2P_{3/2}$) state, each pulse is frequency quadrupled from 906 nm to 226.5 nm (858 nm to 214.5 nm) through phase-matched LBO and BBO nonlinear crystals.  The UV is filtered from the fundamental and second harmonic via dichroic mirrors and directed to the ion with a near transform-limited pulse width of $t_{\text{uv}}\approx1$ ps.  Since the pulsed laser bandwidth ($\sim$400 GHz) is much smaller than the fine-structure splitting ($\sim$74 THz), selective excitation to the different $^2P$ excited states is possible.  Each pulse has $E\approx 10$ pJ of energy, which will excite the ion with a probability of approximately ten percent \cite{excitation:footnote}: $P_{exc}=\text{sin}^2\sqrt{(\gamma^2/4\pi I_s) (Et_{\text{uv}}/w^{2}_{o})}$, where $\gamma$ is the atomic linewidth, $I_s$ is the saturation intensity, and $w_o\approx 6~\mu$m is the beam waist.  This pulsed laser is also used to load ions in the trap via photoionization by tuning to the neutral cadmium $^1S_0$-$^1P_1$ resonance at 228.8 nm.  Once loaded, a single ion will typically remain in the trap for several days.  

After the ion is loaded, it is crystallized within the trap via Doppler cooling on the D2 line at 214.5 nm using the cw laser.  This laser is tuned approximately one linewidth to the red of resonance and localizes the ion to under 1 $\mu$m.  Residual micromotion at the rf drive frequency ($\sim$40 MHz) is reduced via offset electric fields supplied from compensation electrodes \cite{berkland:1998}.  We estimate the kinetic energy from this micromotion to be under 1 Kelvin.

\begin{figure}
\includegraphics[width=1.0\columnwidth,keepaspectratio]{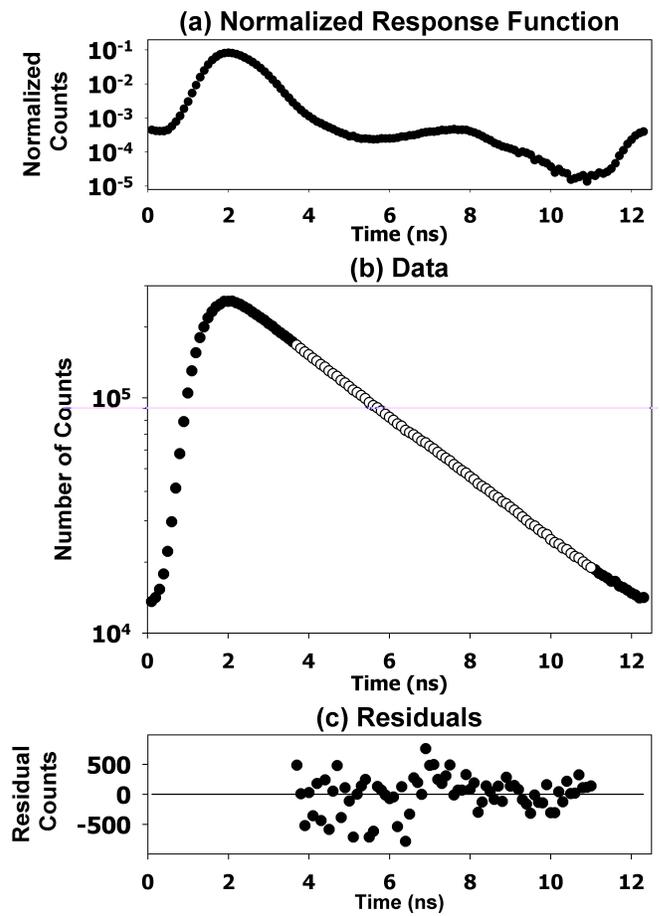}
\caption{\label{fig:data}(a) The response function of the instrument when viewing light scattered off an electrode surface (no atomic physics).  The main peak asymmetry is due to the response time of the PMT of $\approx$ 0.5 ns, whereas the secondary peaks are due to ringing in the PMT electronics ($\sim0.6\%$ of the main peak amplitude).  (b) Data for the $5p$ $^2P_{1/2}$ state taken in the quadrupole trap.  The open circles show the data used to extract the excited state lifetime (see text).  (c) The deviations from the fit function (residuals).}
\end{figure}

Following excitation from the pulsed laser, the spontaneously emitted photons are collected by an $f/2.1$ imaging lens and directed toward a  photon-counting photo multiplier tube (PMT) \cite{aberration:footnote}.  The output signal of the PMT provides the start pulse for the time to digital converter (TDC), whereas the stop pulse is synchronized to the reference clock of the mode-locked laser.  This time-reversed mode is used to eliminate dead time in the TDC.  The PMT used is a Hamamatsu H6240 Series PMT of quantum efficiency $\approx20\%$, and the TDC is an ORTEC model 9353 time digitizer that has 100 ps digital time resolution with no interpolator, accuracy within 20 ppm, less than 145 ps time jitter, and an integral non-linearity within 20 ps rms.

In the experiment, an acousto-optic modulator (AOM) is used to switch on the cw beam to Doppler cool the ion for 500 ns.  Following the cooling pulse, a reference clock from the pulsed laser (synchronized with the laser pulse train) triggers an AOM in the pulsed laser beam and directs a number of pulses to the ion ($\approx$ 15 pulses, with adjacent pulses separated by $\approx$ 12.4 ns).  The repetition rate of this cycle is limited to 1 MHz due to the update time of the pulse generator, and during a given excitation pulse the success probability of detecting an emitted photon is $\sim2\times10^{-4}$.  This gives an average count rate of about 3000 counts per second and thus an expected statistical precision of $\Delta\tau_{rms}/\tau\approx0.25\%/\sqrt{T}$, where $\tau$ is the excited state lifetime and $T$ is data collection time in minutes.

\begin{figure*}
\includegraphics[width=2.0\columnwidth,keepaspectratio]{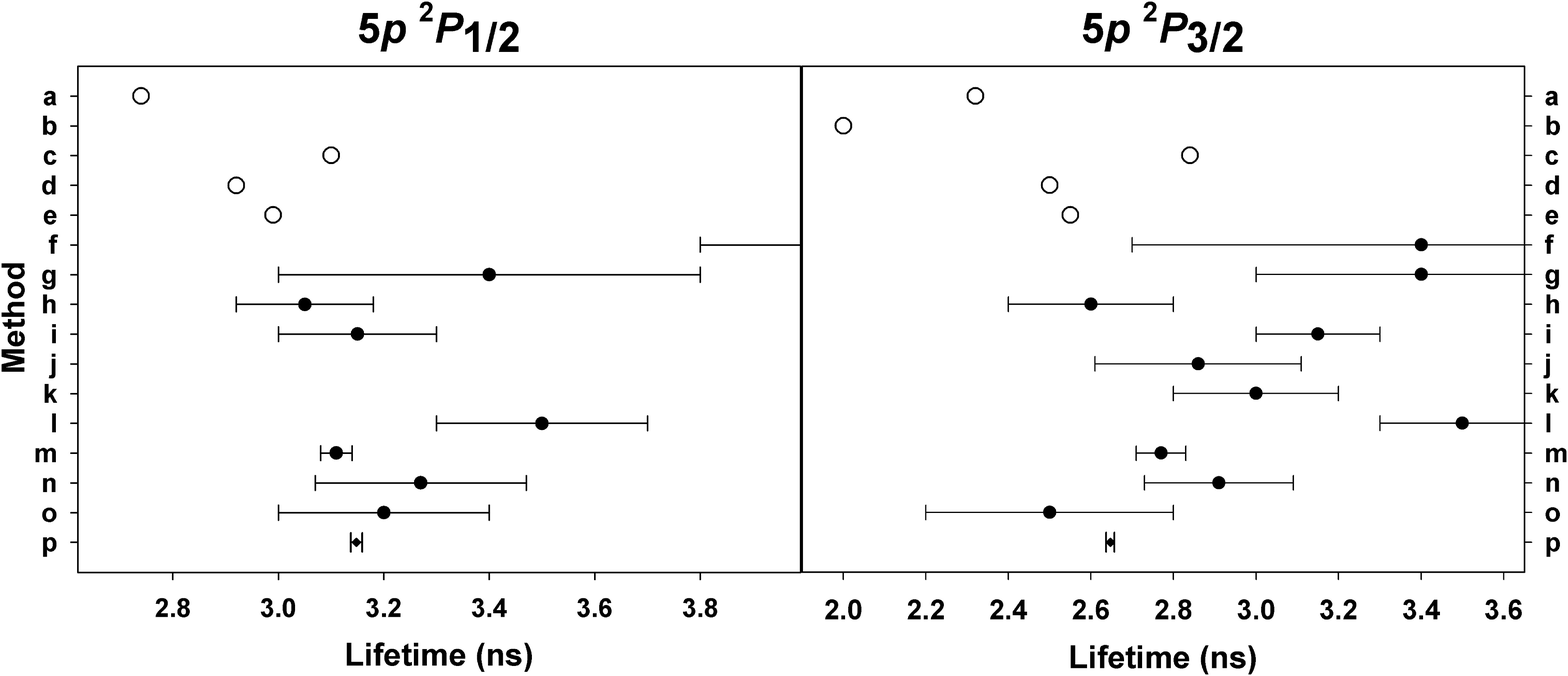}
\caption{\label{fig:P31}Published results of theoretical (open circles) and experimental (filled circles) lifetimes, including this work (filled diamonds), for the $5p$ $^2P_{1/2}$ and $5p$ $^2P_{3/2}$ states of Cd$^+$.
\footnotesize{
(a) Hanle Theory (1974) \cite{hamel:1974},
(b) Theory (1975) \cite{kunisz:1975},
(c) Many Body Perturbation Theory (1997) \cite{chou:1997},
(d-e) Pseudorelativistic Hartree-Fock Theory (2004) \cite{xu:2004},
(f) Phaseshift (1970), $^2P_{1/2}$ value is 4.8 ns \cite{baumann:1970},
(g) Beam-Foil (1973) \cite{andersen:1973},
(h) Hanle (1974) \cite{hamel:1974},
(i) Electron-Photon (1975) \cite{shaw:1975},
(j) Hanle (1976) \cite{rambow:1976},
(k) Hanle (1976) \cite{andersen:1976},
(l) Delayed Coincidence (1980) \cite{verolainen:1980},
(m) Beam-Laser (1994) \cite{pinnington:1994},
(n) Beam-Foil (1994) \cite{pinnington:1994},
(o) Laser-Induced Fluorescence (2004) \cite{xu:2004},
(p) This Experiment.
}}
\end{figure*}

Despite the absence of previously mentioned common systematic effects, possible effects that still must be considered in this system include Zeeman and hyperfine quantum beats \cite{silverman1:1978}.  Zeeman quantum beats have no significant effect (shifts of $<0.05\%$) when working in sufficiently low magnetic fields ($<0.5$ Gauss), whereas hyperfine beating is eliminated by using an even isotope of Cd that has no hyperfine structure (i.e. $^{110}$Cd$^+$).  Potential effects from off-resonant laser light - AC stark shifts, background counts, etc. - are also greatly reduced or eliminated in this experiment by taking data only when the cw cooling beam is switched off via the AOM.  Hence, immediately following the excitation pulse, the only light present is the single spontaneously emitted photon from the ion.  Other possible effects such as relativistic shifts or isotopic dependencies are negligible.  Therefore, the only remaining systematic effects to be considered in this experiment are those arising from the equipment.

\begin{table}
\caption{\label{table}Lifetime measurement results (ns).  The asymmetric quadrupole and linear trap results are in good statistical agreement for the $^2P_{3/2}$ transition and the final result is a weighted average of the two values (the systematic error is common to both).  For the $^2P_{1/2}$ transition, the contribution from the linear trap is omitted from the final result due to an order of magnitude larger prompt peak giving rise to an unusually large systematic error.}
\begin{ruledtabular}
\begin{tabular}{lccr}
Trap & Error & $5p$ $^2P_{1/2}$ & $5p$ $^2P_{3/2}$\\
\hline
Quadrupole & \ldots & 3.148 & 2.646\\
 & Statistical & 0.005 & 0.002\\
 & Systematic & 0.010 & 0.010\\
Linear & \ldots & 3.132 & 2.649\\
 & Statistical & 0.002 & 0.003\\
 & Systematic & 0.030 & 0.010\\
\hline
Final Results &  & 3.148 $\pm$0.011 & 2.647$\pm$0.010\\
\end{tabular}
\end{ruledtabular}
\end{table}

To determine the excited state lifetime, the data in a 12.4~ns range for each laser pulse are summed and time-inverted.  These spectra are corrected for uncorrelated background events and then fit to a single exponential lifetime, $\tau$.  As the start time of the fit is stepped out from the peak \cite{vallery:2003}, the fitted lifetime for the experimental data has a non-statistical variation of 3 to 5 percent.  This systematic effect is expected given the presence of ``prompt'' events from scattered pulsed laser light and from electronic ringing on the PMT seen in Figure ~\ref{fig:data}(a).  The combination of these two effects distorts the spectrum from a pure exponential.  To account for this variation, an appropriately convolved spectrum of the response function [Fig. ~\ref{fig:data}(a)] and $\text{exp}(-t/\tau)$ plus a $\delta$ function prompt peak is generated, fit in the same manner, and compared to the data to extract the lifetime value.  These values are summarized in Table \ref{table} for each trap and transition.

The final values, summarized in Table \ref{table} for each trap, are $3.148\pm0.011$ ns for the $^2P_{1/2}$ state and $2.647\pm0.010$ ns for the $^2P_{3/2}$ state.  The final error is the average of the statistical error (less than 0.15\% for all measurements) and the systematic error.  The systematic error of approximately 0.4\% is due to the uncertainty in comparison of the fitted values of the convolved spectrum and the experimental data.  These new results are plotted in Figure~\ref{fig:P31} along with previously reported theoretical and experimental values for these levels.  It is seen from this figure that the results reported in this paper are the most precise measurements of these particular excited states of Cd$^+$.

In conclusion, we have demonstrated a new technique for measuring excited state atomic lifetimes that is able to eliminate common systematic errors associated with such measurements.  The results herein are not only the most precise to date for Cd$^+$, but with absolute uncertainties of order 10~ps, are among the most precisely measured excited state lifetimes in any atomic system.  Furthermore, this technique has the potential to achieve $\sim$100 ppm precision by eliminating the remaining systematic effects due to prompt events and PMT ringing.  Other possible improvements include increasing the data collection rate by using a faster pulse generator and TDC, and measuring a longer decay range by pulse-picking individual pulses.

\begin{acknowledgments}
This work is supported by the U.S. National Security Agency and Advanced Research and Development Activity under Army Research Office contract DAAD19-01-1-0667 and the National Science Foundation Information Technology Research Program. 
\end{acknowledgments}


\begin{thebibliography}{32}
\bibitem[{exc()}]{excitation:footnote}
\bibinfo{note}{It would be possible to increase the data rate by increasing
  this excitation probability, however, it was kept to near 10\% since higher
  pulsed laser powers also tend to load extra ions into the trap.}
\bibitem[{abe()}]{aberration:footnote}
\bibinfo{note}{Due to the chromatic aberration of this imaging system,
  state-selective light collection between $^2P_{1/2}$ and $^2P_{3/2}$ is also
  achieved.}
\end{thebibliography}

\end{document}